# Discovery of Weyl semimetal state violating Lorentz invariance in MoTe$_2$

N. Xu[1,2,*,§], Z. J. Wang[3,*], A. P. Weber[1,2,*], A. Magrez[1], P. Bugnon[1], H. Berger[1], C. E. Matt[2,7], J. Z. Ma[4], B. B. Fu[2,4], B. Q. Lv[2,4], N. C. Plumb[2], M. Radovic[2], E. Pomjakushina[5], K. Conder[5], T. Qian[4], J. H. Dil[1,2], J. Mesot[1,2,6], H. Ding[4,7,§] and M. Shi[2,§]

[1] Institute of Physics, École Polytechnique Fédérale de Lausanne, CH-10 15 Lausanne, Switzerland

[2] Swiss Light Source, Paul Scherrer Institut, CH-5232 Villigen PSI, Switzerland

[3] Department of Physics, Princeton University, Princeton, NJ 08544, USA

[4] Beijing National Laboratory for Condensed Matter Physics and Institute of Physics, Chinese Academy of Sciences, Beijing 100190, China

[5] Laboratory for Developments and Methods, Paul Scherrer Institut, CH-5232 Villigen, Switzerland

[6] Laboratory for Solid State Physics, ETH Zürich, CH-8093 Zürich, Switzerland

[7] Collaborative Innovation Center of Quantum Matter, Beijing, China

* These authors contributed equally to this work.

§ E-mail: nan.xu@psi.ch, dingh@iphy.ac.cn, ming.shi@psi.ch

A new type of Weyl semimetal state, in which the energy values of Weyl nodes are not the local extrema, has been theoretically proposed recently, namely type-II Weyl semimetal. Distinguished from type-I semimetal (e.g. TaAs), the Fermi surfaces in a type-II Weyl semimetal consist of a pair of electron- and hole- pockets touching at the Weyl node. In addition, Weyl fermions in type-II Weyl semimetals violate Lorentz invariance. Due to these qualitative differences distinct spectroscopy and magneto-transport properties are expected in type-II Weyl semimetals. Here, we present the direct observation of the Fermi arc states in $MoTe_2$ by using angle-resolved photoemission spectroscopy. Two arc states are identified for each pair of Weyl nodes whoes surface projections of them possess single topological charge, which is a unique property for type-II Weyl semimetals. The experimentally determined Fermi arcs are consistent with our first-principle calculations. Our results unambiguously establish that $MoTe_2$ is a type-II Weyl semimetal, which serves as a great test-bed to investigate the phenomena of new type of Weyl fermions with Lorentz invariance violated.

The studies of topological Weyl semimetals (WSM) provide a new frontier in the field of topological quantum states [1-13]. A WSM possesses unique band structure with pairs of crossing points (called Weyl nodes) near the Fermi level ($E_F$) in the bulk, similar to those of Dirac semimetals [14-19], but the spin degeneracy of the energy bands is lifted by breaking of time reversal symmetry or inversion symmetry. A Weyl node is a sink or a source of Berry curvature, which can be viewed as a magnetic monopole in the momentum space ($k$-space) [20], and the low-energy excitations near Weyl nodes behave the same as Weyl fermions. The theoretically predicted Weyl nodes [21-22] have been directly observed in TaAs family [23-24] (illustration shown in Fig. 1a,b), and key signatures of Weyl fermions such as surface Fermi arcs and chiral anomaly induced negative magneto-resistance have also been unveiled by spectroscopy [25-36] and magneto-transport measurements [37-40], respectively.

A new type of WSM states (namely type-II WSM) has been brought forward by theoretical consideration [41]. Different from the standard WSM states in TaAs family, the energy values of Weyl nodes in type-II WSM are not the local extrema, as illustrated in Fig. 1c,d. Fermi surfaces (FS) in type-II WSM consist of pairs of electron- and hole- pockets touching at the Weyl node. In addition, the Weyl fermion quasi-particles violate Lorentz invariance in type-II WSM. Due to these qualitative differences from the standard WSM states in TaAs family, distinct spectroscopy and magneto-transport properties are expected in this new type of WSM. Type-II WSM has been theoretically proposed in $WTe_2$ [41] and $(W,Mo)Te_2$ [42], however, the direct evidence remain experimentally elusive because the Weyl nodes are close to each other and the related features are too dim to be observed [43].

A recent theoretical work [44] suggests that the orthorhombic ($T_d$) phase of $MoTe_2$ ($MoTe_2$ discussed in the manuscript is only for this structural polytype), which is iso-structural with $WTe_2$ (Fig. 1e), is a type-II WSM candidate with 8 well-separated Weyl nodes forming 2 groups in different energy positions. Later calculation with lattice parameters determined from experiments at lower temperature (100 K), indicates a different topological physics in $MoTe_2$ in which only 4 Weyl nodes exist near $E_F$ in the $k_z = 0$ plane of the bulk Brillouin zone (BZ), as illustrated in Fig. 1f,h [45]. This makes $MoTe_2$ possibly to be the simplest type-II WSM in the presence of time reversal symmetry, with the minimal possible number (4) of Weyl nodes in the BZ. In this case, Fermi arc states, which are well separated from bulk states, appear below $E_F$, which are possible to be visualized by angle-resolved photoemission spectroscopy (ARPES). Moreover, other quantum phases like topological nodes-line semimetal state [46-53] and superconductivity have also been found in MoTe$_2$ recently [45,54]. Characterization of the bulk band structure and possible topological surface states is a critical step toward understanding the diverse topological physics in MoTe$_2$.

Here we present a comprehensive ARPES investigation on high quality single crystal MoTe$_2$, combined with first-principle calculations. Surface states are identified from the bulk bands by photon energy dependent measurements. By counting the crossing points through a closed loop in $k$-space, we identified the topological Fermi arc states induced by the type-II Weyl nodes, which is fully consistent with our first principle calculations. We observed two arcs connect each pair of type-II Weyl nodes with single topological charge (±1), which is the unique feature of the type-II WSM. Our results provide compelling evidence that MoTe$_2$ is a type-II Weyl semimetal with the Lorentz invariance violated, which serves as a great test-bed to investigate the phenomena of new type of Weyl fermions with Lorentz invariance violated.

Figure 2a shows the band dispersions along the high symmetry line $\bar{\Gamma}$-$\bar{X}$, obtained from the ARPES measurements on the (001) surface of MoTe$_2$ using hν = 20 eV photons, which corresponding to $k_z$ = 0 plane. The well-defined quasi-particles are further observed in the plot of energy distribution curves (EDC) plot in Fig. 2b. The overall feature of the measured band structure is in a good agreement with the calculated bulk states along $\Gamma$-$X$ (Fig. 2c), indicating its possible bulk origination. To further examine the dimensionality of the measured bands, we performed photon energy dependent ARPES measurements. As seen from FS map in the $k_x$-$k_z$ plane shown in Fig. 2e-f, both the intensity and the Fermi momentum vary with $k_z$, which confirms the bulk originations of the bands crossing $E_F$ along $\bar{\Gamma}$-$\bar{X}$. On the other hand, another band is observed around 50 meV below $E_F$ (arrow in Fig. 2a), which is absent in the calculated bulk electronic structure (Fig. 2c), suggesting that it may have different origin than the bulk states. This band appears in our calculated spectral function of the surface states (pointed by the arrow in Fig. 2d), which is sharp and well separated from other blur features that are the projection of bulk states on surface. Our theoretical analysis indicates that this band is the topological surface state (TSS) guaranteed by the non-trivial Z$_2$ invariant of the $k_y$ = 0 plane [45], which will be discussed more in the later part of this paper. As shown from the curvature plots in Fig. 2g-j, the location of this arc band does not change when varying the incident photon energy in a wide range (corresponding $k_z$ value changes around $4\pi/c$), which indicates that this is a surface state, consistent with our surface calculations.

To investigate how the TSS disperses in the surface Brillouin zone, we present the FS map in Fig. 3a. The measured FS shows remarkable agreements with that from our calculations (Fig. 2b). In total we observed four bulk pockets sitting along the $\bar{\Gamma}$-$\bar{X}$ direction, which can be resolved in the curvature plot near $E_F$ along cut 1 (Fig. 3c). One hole-like band forms a more or less circle-shape pocket (h1) around the $\bar{\Gamma}$ point, together with the other hole-like band with a cross-shape (h2). Two electron-like pockets are sitting both sides of the $\bar{\Gamma}$ point. The intense one, which is farther from the center, is in a crescent-shape (e1) and the dim one is closer to the $\bar{\Gamma}$ point in a strawberry-shape (e2). These two electron pockets are formed by the same bulk energy band, as indicated by our calculation in Fig. 2c. Distinct from other pockets, the TSS doesn't cross $E_F$ and merge into the hole pocket below $E_F$ in $\bar{\Gamma}$-$\bar{X}$ high symmetry line, which is consist with our calculations in Fig. 3d (more details in Ref. [45]). Away from the $\bar{\Gamma}$-$\bar{X}$ mirror plane, the TSS crosses $E_F$ as indicated by our band structure calculation (Fig. 3h), which can be clearly resolved in the ARPES intensity plot along cut 2 (Fig. 3f), the corresponding curvature (Fig. 3g) and EDC plot (Fig. 3i). Further tracing this band by acquiring ARPES spectra in more cuts, we determine that the FS of this state forms two discrete arcs aside the $k_y = 0$ mirror plane, connecting the bulk electron- and hole-pockets. The FS topology determined by our measurements is illustrated in Fig. 3e.

A common practice for demonstrating the existence of the topological Fermi arc states in WSM is to count how many times a closed loop intersects with the FS crossing in $k$-space [25,33]. An odd number of FS crossings are the compelling evidence for Fermi arcs, because any closed Fermi pockets can only cross a loop an even number of times. The loop chosen here is indicated in Fig 3a. Along L1, which is part of the $\bar{\Gamma}$-$\bar{X}$ line, there are three crossing points as seen from Fig. 3c: e1 crosses twice and e2 crosses once. Along L2 (part of the cut 2 line in Fig. 3f,g), there are four crossing points in total: e2 crosses once, h2 crosses twice and the arc state crosses once. Along L3 and L4, there is no crossing point (Fig. 3j,k). The crossing points along the chosen loop in momentum space are indicated as red points in Fig. 3e. All together, we determined that the FSs cross the closed loop seven times, thus providing direct experimental evidence of the existence of Fermi arcs on the (001) surface of $MoTe_2$. The observed dispersion of the arc states, which connect electron pocket e1 and hole pocket h2, is fully consistent with our first-principle calculations (Fig. 3b). More detailed theoretical discussion on the assignment of topological Fermi arc states in $MoTe_2$ can be found in Ref. [45].

In the TaAs family type-I WSM, the Fermi arc states formed by the TSS are observed in a certain energy range near the Weyl nodes [23-36]. As shown in the panel I of Fig. 4a, when the chemical potential is below the Weyl nodes energy, the Fermi arc states in type-I WSM end at pairs of bulk hole pockets which are the lower part of the bulk Weyl cones (Fig. 1a). Rising the chemical potential to Weyl nodes energy, the bulk hole-pockets shrink to points (Weyl nodes) and Fermi arcs connect Weyl nodes with different topological charge (panel II in Fig. 4a). Further lifting the chemical potential above the nodes energy, the Weyl nodes expand to pairs of electron-pockets, and the Fermi arcs still present between them (panel III in Fig. 4a). Similarly, in the type-II WSM like in $MoTe_2$, the Fermi arcs also disperse in a certain energy window. For this reason, we can direct observe the unclosed arc states although the Weyl nodes locates ~80 meV above $E_F$ in our sample. On the other hand, the evolution of Fermi arc states with energy is quite different in type-II WSM such as $MoTe_2$, as depicted in Fig. 4b. Below the Weyl nodes energy (panel I of Fig. 4b), the bulk electron pockets (blue circles) and hole pockets (red circles) are well separated from each other. The Fermi arcs are already present at this energy: only one arc for each pair of type-II Weyl nodes structure with both ends terminated at the bulk electron pockets. Moving the chemical potential more close to the node energy (panel II in Fig. b), although the bulk electron and hole pockets are not touching yet, guided by the nontrivial $Z_2$ invariant of the vertical ($k'' = 0$) plane, (the same as the $k_y = 0$ plane in $MoTe_2$), the middle part of each arc should merge into the bulk hole pocket. As a result, the arc is cut into two pieces which connect the bulk electron and hole pockets. The Fermi arc state of $MoTe_2$ presented in this paper is in this case. Further rising the chemical potential, the arcs shrink and finally disappear when the bulk electron and hole pockets touching at the Weyl nodes energy (panel III in Fig. 4b). The double-arcs structure connecting a pair of Weyl nodes with single topological charge (±1) are the unique feature of type-II WSM, and is observed in $MoTe_2$ for the first time.

In summary, using ARPES we have observed Fermi arcs on the (001) surface of MoTe$_2$. Along a closed loop in the *k*-space, an odd number of FS crossing points is determined, which is a sufficient condition for the existence of Fermi arcs. The experimentally determined surface states are in a remarkable agreement with our first-principles calculations. The double arcs structure associated with one pair of Weyl nodes with a single topological charge, which is unique property of type-II WSM, is unveiled in our ARPES measurements. The first observation of the Fermi arcs in MoTe$_2$, together with the first-principle calculations, established the discovery of type-II WSM in MoTe$_2$.

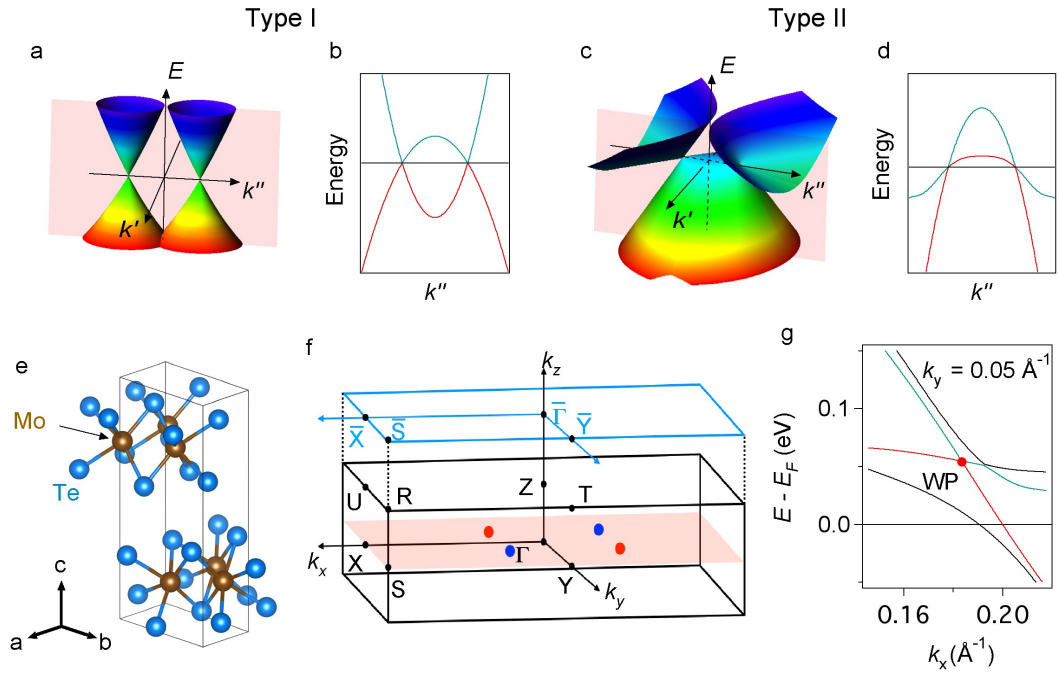

**Fig. 1. The prediction of type-II Weyl semimetal states in MoTe$_2$. a,** Illustration of band dispersions of a pair of type-I WSM in the two-dimensional plane containing nodes. **b,** The Schematic of band dispersions passing through a pair of type-II Weyl nodes, as indicated by the red plane in **a**. **c-d,** Same as **a-b**, but for type-II WSM. **e,** Crystal structure of orthorhombic (T$_d$) phase of MoTe$_2$. **e,** Surface and bulk Brillouin zone of MoTe$_2$. High-symmetry points are labelled. **g,** Type-II Weyl nodes in MoTe$_2$ obtained from the first-principles calculations.

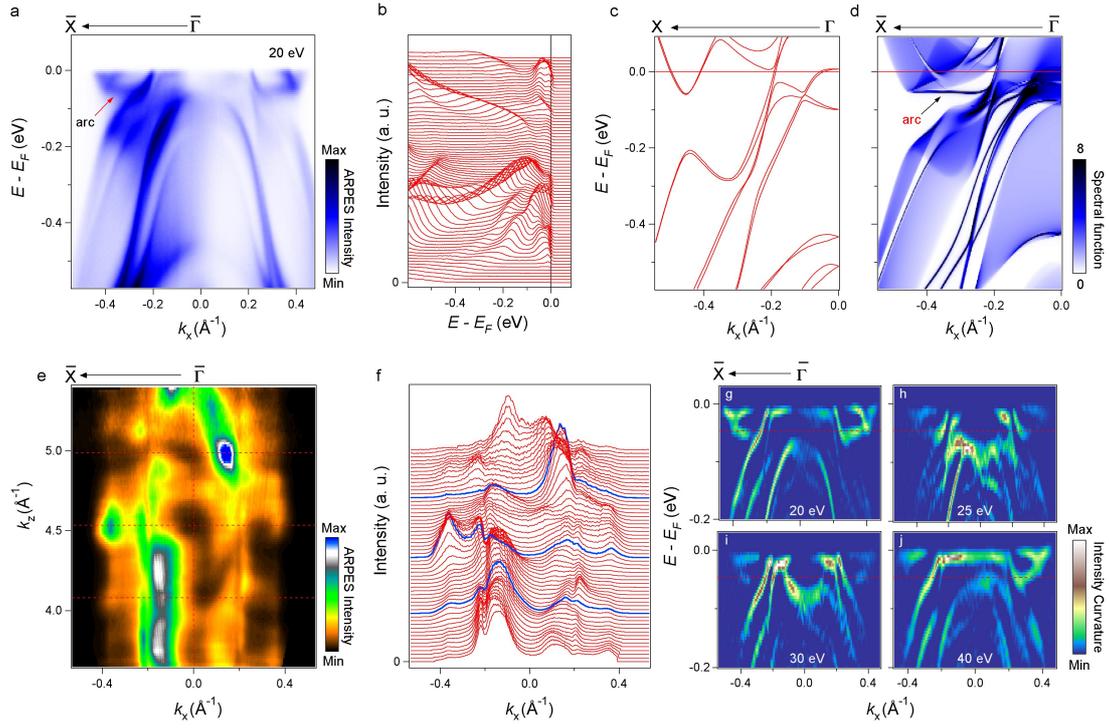

**Fig. 2. The band structure of MoTe$_2$ obtained from ARPES measurements and first-principle calculations. a-b,** ARPES intensity and EDC plots along the $\bar{\Gamma}$-$\bar{X}$ direction, taken with hν = 20 eV. **c,** The calculated bulk band structure along the $\Gamma$-$X$ direction. **d,** The calculated spectral function of surface states. **e-f,** The FS intensity map in the $k_x$-$k_z$ plane and its corresponding EDC plot. **g-j,** ARPES intensity curvature plot along the $\bar{\Gamma}$-$\bar{X}$ direction, taken with different incident photon energies, which covers a $k_z$ range ~ 4π/c.

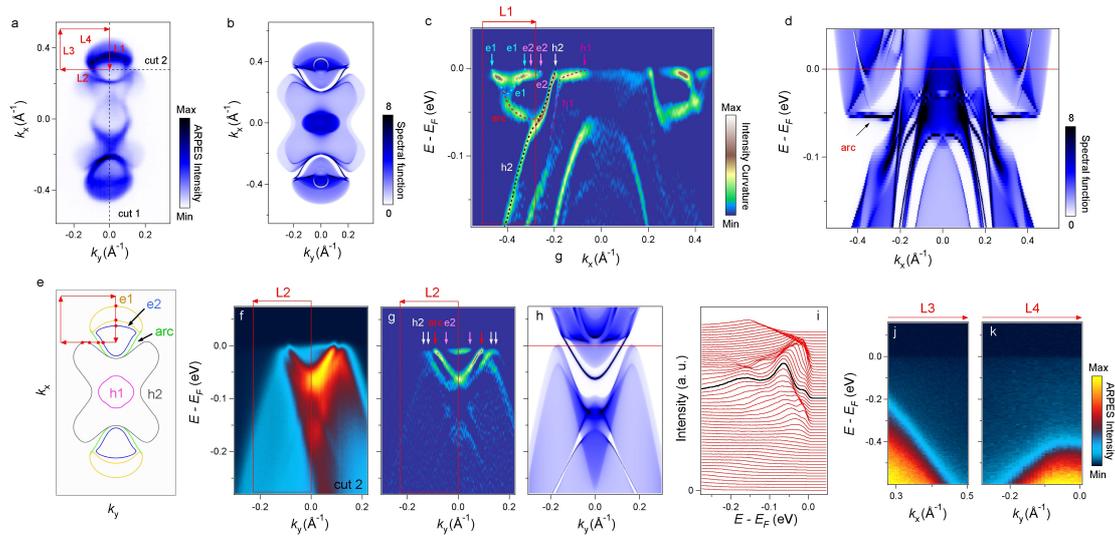

**Fig. 3. Fermi arc states in MoTe$_2$. a,** FS intensity map in the $k_x$-$k_z$ plane, obtained from ARPES measurements with hν = 20 eV. **b,** The spectral function from the surface states calculation. **c-d,** ARPES intensity curvature plot near $E_F$ along the $\bar{\Gamma}$-$\bar{X}$ direction and the corresponding spectral function from the surface states calculation. **e,** The Schematic of FS topology in MoTe$_2$. The closed loop and crossing points for detecting Fermi arc states are indicated. **f,** The ARPES intensity along cut 2 in **a**. **g-i,** The curvature plot of the ARPES intensity, the spectra function from the surface states calculation and EDC plots of the spectrum in **f**, respectively. **j-k,** ARPES intensity plot along L3 and L4, which are part of the closed loop.

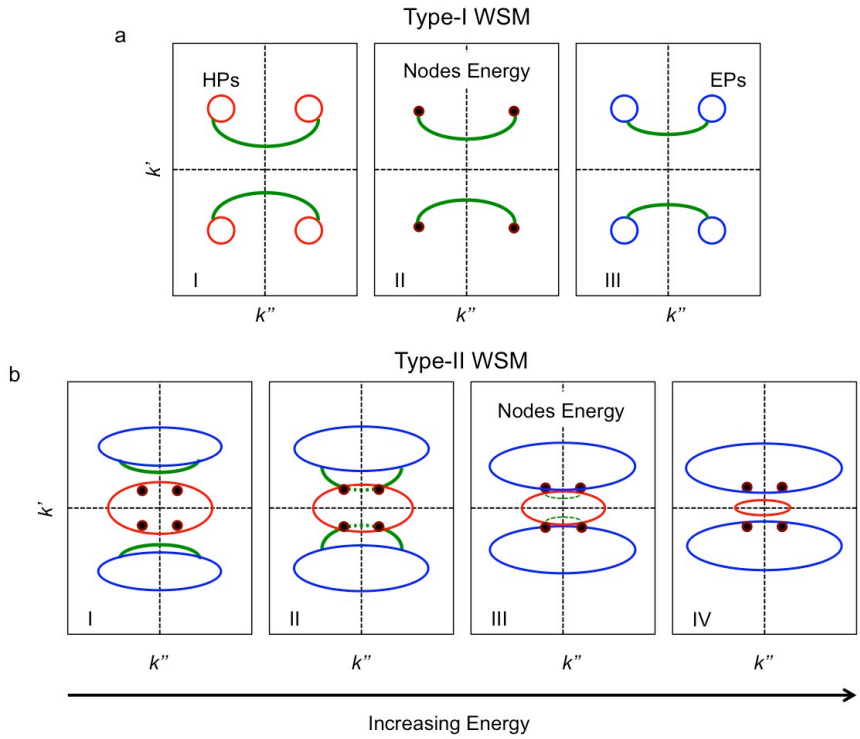

**Fig. 4. Evolution of Fermi arc states near the Weyl nodes as a function of energy. a-b,** The Schematic of Fermi arc state evolution in type-I and type-II WSM, respectively. The bulk electron-pockets (EPs) and bulk hole pockets (HPs), which form the Weyl nodes at their touching points, are plotted in blue and red circles, respectively. The black dots present the Weyl nodes. The arcs are indicated as green solid lines. The green dashed lines indicate the part which merged into the bulk states with hybridization.